\documentclass[prd,preprint,
superscriptaddress,showpacs,nofootinbib,%
tightenlines,%twocolumn
]{revtex4}
\usepackage{epsfig}
\usepackage{slashed}
\usepackage{amssymb}

\newcommand{\ben}{\begin{displaymath}}
\newcommand{\een}{\end{displaymath}}
\newcommand{\be}{\begin{equation}}
\newcommand{\ee}{\end{equation}}
\newcommand{\bea}{\begin{eqnarray}}
\newcommand{\eea}{\end{eqnarray}}
\begin{document}
%\preprint{MKPH-T-10-}
\title{Path integral quantization for massive vector bosons}
\author{D.~Djukanovic}
\affiliation{Institut f\"ur Kernphysik, Johannes
Gutenberg-Universit\"at, D-55099 Mainz, Germany}
\author{J.~Gegelia}
\affiliation{Institut f\"ur Kernphysik, Johannes
Gutenberg-Universit\"at, D-55099 Mainz, Germany} \affiliation{High
Energy Physics Institute, Tbilisi State University, 0186 Tbilisi,
Georgia}
\author{S.~Scherer}
\affiliation{Institut f\"ur Kernphysik, Johannes
Gutenberg-Universit\"at, D-55099 Mainz, Germany}
\date{January 7, 2010}
\begin{abstract}
   A parity-conserving and Lorentz-invariant effective field theory of
self-interacting massive vector fields is considered. For the
interaction terms with dimensionless coupling constants the
canonical quantization is performed.
   It is shown that the self-consistency condition of this system with the second-class
constraints in combination with the perturbative renormalizability
leads to an SU(2) Yang-Mills theory with an additional mass term.
\end{abstract}

%\pacs{
%03.65.Nk,
%Nonrelativistic scattering theory
%11.10.Gh,
%Renormalization
%12.39.Fe.}
%Chiral Lagrangians
%04.60.Ds  Canonical quantization
%04.60.Gw  Covariant and sum-over-histories quantization
%03.70.+k  Theory of quantized fields
%          (see also 11.10 Field theory)


\maketitle

\section{Introduction}

   Vector mesons play an important role in the phenomenological
description of hadronic processes (see, for example,
Ref.~\cite{Drechsel:1998hk}).
   Phenomenological low-energy chiral Lagrangians with vector mesons
were already constructed in the 1960s
\cite{Schwinger:1967tc,Wess:1967jq,Weinberg:de}.
   Details of the construction of chirally invariant effective
Lagrangians describing the interaction of vector mesons with
pseudoscalars and baryons can be found, e.g., in
Refs.~\cite{Gasser:1984yg,Meissner:1987ge,Bando:1988br,
Ecker:yg,Ecker:1988te,Borasoy:1996ds,Birse:1996hd,Harada:2003jx}.
   Besides a knowledge of the most general effective
Lagrangian, an effective field theory (EFT) program requires a
systematic power counting.
   In Refs.\ \cite{Fuchs:2003sh,Schindler:2003xv} it was shown how
to consistently include virtual (axial-) vector mesons in EFT, if
they appear only as internal lines in Feynman diagrams involving
soft external pions and nucleons with small three-momenta.
   This is possible if a suitable renormalization condition such as the
extended on-mass-shell renormalization scheme or the infrared
renormalization is chosen (see also
Refs.~\cite{Bruns:2004tj,Bruns:2008ub}).
   In this case the masses of the vector mesons are treated as a large scale.
   For energies for which the vector mesons can be generated, the problem of
the power counting is solved by using the complex-mass
renormalization scheme
\cite{Stuart:1990,Denner:1999gp,Denner:2006ic,Denner:2005fg,Actis:2006rc,Actis:2008uh},
which is an extension of the on-mass-shell renormalization scheme
for unstable particles \cite{Djukanovic:2009zn}. Finally, it would
also be possible to consider an EFT with a heavy scale much larger
than the masses of the vector bosons. For example, if the Higgs
boson were never discovered then such an EFT would appear as a
possible candidate for a theory of the (electro-)weak interaction.

A covariant Lagrangian formalism for massive vector fields entails
a special feature: they are described by Lagrangians with
constraints.
   The self consistency of a system with constraints imposes conditions
on the form of the Lagrangian.

    In Ref.~\cite{Djukanovic:2004mm} the effective Lagrangian of
Ref.~\cite{Weinberg:de} describing the interaction among $\rho$
mesons, pions, and nucleons was considered. Requiring the
perturbative renormalizability in the sense of effective field
theory \cite{Weinberg:mt}, the universality of the vector-meson
coupling was derived. It has been argued \cite{Birse:1996hd,f2}
that the effective Lagrangian of Ref.~\cite{Weinberg:de} is not
the most general one because it starts with the Yang-Mills
structure for the self-interacting part of the vector mesons and
takes the coupling to the fermion doublet equal to the coupling of
the vector-meson self interaction.

   In this work we will close this gap.
   We start with the most general effective Lagrangian of a system of three
interacting massive vector fields.
   In principle, the Lagrangian contains {\em all} interaction terms which
respect Lorentz invariance, hermiticity, parity, and charge
conservation.
   Assuming that the interaction terms with a higher number of derivatives and/or
fields are suppressed by higher orders of some large scale we
consider only interactions with dimensionless coupling constants.
   We analyze the quantization in the Hamilton formalism.
   To have a constrained system with the right number of degrees
of freedom the determinant of the matrix of constraints should be
non-vanishing.
   This nontrivial condition leads to relations among the
coupling constants of the Lagrangian.
   Furthermore, demanding the perturbative renormalizability in the sense of
EFT we arrive at additional consistency conditions for the parameters of the
Lagrangian.
   As will be shown below, we are eventually led to the Yang-Mills structure
incorporated in the effective Lagrangian of Ref.~\cite{Weinberg:de}.

\section{Lagrangian}

   We consider a system of three interacting massive vector fields
respecting Lorentz invariance and parity conservation.
   We assume that two vector fields correspond to a pair of
charged particles, $V^\pm_\mu=(V^1_\mu\mp iV^2_\mu)/\sqrt{2}$, while
the third component, $V^3_\mu$, is neutral. The most general
effective Lagrangian respecting all given symmetries contains an
infinite number of terms and can be split as
   \begin{equation}
{\cal L}_{V\rm eff}={\cal L}_{A}+{\cal L}_{B}+{\cal L}_{C}.
\end{equation}
   Here, ${\cal L}_{A}$ refers to the free Lagrangian plus a finite number
of interaction terms with dimensionless coupling constants,
${\cal L}_{B}$ contains an infinite number of ``non-renormalizable''
interactions, and ${\cal L}_{C}$ refers to interactions with other
degrees of freedom.
   In this work we only treat the vector-meson self-interaction terms
with dimensionless coupling constants, assuming that the ``non-renormalizable''
interactions are suppressed by powers of some large scale,
\begin{equation}
{\cal L}_A= {\cal L}_2+{\cal L}_3+{\cal L}_4\,.\label{LagGeneral}
\end{equation}
The free Lagrangian reads
\begin{equation}
{\cal L}_2  = -{1\over 4} \ V^a_{\mu\nu} V^{a \mu\nu}
+\frac{M_a^2}{2} V_\mu^a V^{a \mu}\,. \label{kvLagr}
\end{equation}
   Here, $V^a_{\mu\nu}=\partial_\mu V^a_\nu -
\partial_\nu V^a_\mu $, $M_a$ is the mass of $a$-th vector field
($M_1=M_2=M$),
and a summation over $a$ from 1 to 3 is implied.
   The interaction terms with dimensionless coupling constants involve either
three or four vector fields.
   The interaction Lagrangian with three vector fields
is of the form \cite{3vc}
\begin{equation}
{\cal L}_3 = - g^{abc} V^a_\mu V^b_\nu \partial^\mu V^{c \nu}\,,
\label{3VintLagr}
\end{equation}
where $g^{abc}$ ($a,b,c=1,2,3$) are coupling constants.

We use charge conservation and hermiticity to restrict the coupling
constants $g^{abc}$.
   Hermiticity requires that all of these
couplings are real.
   The invariance of the Lagrangian of
Eq.~(\ref{3VintLagr}) under a global U(1) transformation with an
infinitesimal parameter $\varepsilon$,
\begin{equation}
V^a_\mu \rightarrow V^a_\mu-\varepsilon\, \epsilon^{3ab} V_\mu^b\,,
\label{transfrules}
\end{equation}
leads to
\begin{equation}
(g^{abd}\epsilon^{3dc}+g^{adc}\epsilon^{3db}+g^{dbc}\epsilon^{3da})V^a_\mu
V_\nu^b
\partial^\mu V^{c\nu}=0\,. \label{grelations}
\end{equation}
   Using Eq.~(\ref{grelations}) we express  the couplings in terms of
seven real parameters,
\begin{eqnarray}
g^{333} & = & g_1, \ \ \ g^{113}= g_2, \ \ \ g^{123}= -g_3, \ \
\ g^{213}= g_3,\nonumber\\
g^{223} & = & g_2, \ \ \ g^{311}= g_4, \ \ \ g^{321}= -g_5, \ \ \
g^{312}= g_5, \nonumber\\
g^{322} & = & g_4, \ \ \ g^{131}= g_6, \ \ \ g^{231}= -g_7, \ \ \
g^{132}= g_7,\nonumber\\ g^{232} & = & g_6\,. \label{3couplings}
\end{eqnarray}
All other constants vanish.
   Note that at this stage we did not demand invariance
under a charge-conjugation transformation (C), $V_\mu^a\mapsto (-)^a
V_\mu^a$, which would, in addition, eliminate all constants with
$a+b+c$ odd, namely $g_1$, $g_2$, $g_4$, and $g_6$.

   The interaction Lagrangian involving four vector fields has the form
\begin{equation}
{\cal L}_4 = - h^{a b c d} V_\mu^a V_\nu^b V^{c \mu}V^{d\nu}\,,
\label{4VL}
\end{equation}
where $h^{a b c d}$ ($a,b,c,d=1,2,3$) are real coupling constants
satisfying the permutation symmetries
\begin{eqnarray}
\label{gabcdsym}
h^{abcd}&=&h^{bcda}=h^{cdab}=h^{dabc}\nonumber\\
&=&h^{cbad}=h^{adcb}=h^{dcba}=h^{badc}.
\end{eqnarray}
   Invariance under the U(1) transformation
leads to
\begin{equation}
(h^{ebcd}\epsilon^{3ea}+h^{aecd}\epsilon^{3eb}+h^{abed}\epsilon^{3ec}
+h^{abce}\epsilon^{3ed})V_\mu^a V_\nu^b V^{c \mu}V^{d\nu}=0\,.
\label{gabcdrelations}
\end{equation}
   Combining Eqs.\ (\ref{gabcdsym}) and (\ref{gabcdrelations})
allows us to write the couplings $g^{abcd}$ in terms of five real
parameters,
\begin{eqnarray} h^{1111} & = & \frac{d_1+d_2}{4}\,, \nonumber\\
h^{1122} & = & h^{2112}=h^{1221}=h^{2211}=\frac{d_2}{4}\,,\nonumber\\
h^{1313} & = & h^{3131}=\frac{d_4}{4}\,, \nonumber\\
h^{1133} & = & h^{3113}=h^{1331}=h^{3311}=\frac{d_3}{8}\,,\nonumber\\
h^{2121} & = & h^{1212}=\frac{d_1-d_2}{4}\,, \nonumber\\
h^{2233} & = & h^{3223}=h^{2332}=h^{3322}=\frac{d_3}{8}\,,\nonumber\\
h^{2222} & = & \frac{d_1+d_2}{4}\,, \nonumber\\
h^{2323} & = & h^{3232}=\frac{d_4}{4}\,,\nonumber\\
h^{3333} & = & d_5\,. \label{4couplings}
\end{eqnarray}
   All other coupling constants vanish.
   In this case, charge conservation also implies charge conjugation
invariance, i.e., $a+b+c+d$ even.

\section{The Hamiltonian method}
   To quantize the above theory of massive
vector fields we use the canonical formalism following
Ref.~\cite{gitman}.
   The momenta conjugated to the fields $V^a_0$ and $V^a_i$ are defined as
\begin{eqnarray}
\pi^a_0 &=& {\partial {\cal L}_V\over \partial\dot V^a_0}=-g^{bca}
V^b_0 V^c_0\, \label{pi0},
\\
\pi^a_i &=& {\partial {\cal L}_V\over \partial\dot V^a_i}=V^a_{0i}
+g^{bca} V^b_0 V^c_i\,. \label{pii}
\end{eqnarray}
   Note that we will not use a fully covariant notation in deriving the
Hamiltonian.
   The velocities $\dot V^a_0 $ cannot be solved from Eq.\
(\ref{pi0}), i.e.~the corresponding momenta $\pi^a_0$ need to
satisfy the primary constraints
\begin{equation}
  \phi_1^a=\pi^a_0+g^{bca} V^b_0 V^c_0\approx 0. \label{phi1}
\end{equation}
   Here, $\phi^a_1\approx 0$ denotes a weak equation in Dirac's
sense, namely that one must not use one of these constraints before
working out a Poisson bracket \cite{Dirac}.
   On the other hand, from Eq.~(\ref{pii}) we solve
\begin{equation}
\dot V_i^a=\pi_i^a+\partial_iV_0^a-g^{bca}V_0^bV_i^c. \label{aidot}
\end{equation}
   Next we construct the so-called total Hamiltonian density:
\begin{equation}
{\cal H}_1= \phi^a_1 z^a +{\cal H}\,, \label{hamden}
\end{equation} where
\begin{eqnarray}
{\cal H} & = & {\pi^{a}_i\pi^{a}_i\over 2} +\pi_i^a
\partial_i V^a_0 + \frac{1}{4} V^a_{ij}
V^a_{ij} - {M_a^2\over 2}V^a_\mu V^{a \mu}\nonumber\\
& &-g^{abc}V^a_0 V^b_i \pi^c_i- g^{abc} V_0^a V_i^b
\partial_iV^c_0-g^{abc} V_i^a V_0^b
\partial_iV^c_0
\nonumber\\
& & + g^{abc} V_i^a V_j^b \partial_i V^c_j+\frac{1}{2} g^{abc}
g^{a'b'c} V_0^a V_i^b
V^{a'}_0 V_i^{b'} \nonumber\\
& &+ h^{abcd} V_\mu^a V_\nu^b V^{c \mu} V^{d \nu}\,.
\label{hamiltonian}
\end{eqnarray}
   In Eq.\ (\ref{hamden}) $z^a$ are arbitrary functions which have to
be determined.

   The primary constraints have to be conserved in time. Therefore, for
each $a$, we calculate the Poisson bracket of $\phi^a_1$ with the
Hamiltonian
 \be H_1=\int d^3{\bf x}\, {\cal
H}_1({\bf x})\,, \label{truehamiltonian} \ee and obtain
\begin{eqnarray}
\left\{ \phi^a_1,H_1\right\}&=&
\left(g^{bca}+g^{cba}-g^{acb}-g^{cab}\right) V_0^c z^b \nonumber\\
&  &+ \partial_i\pi_i^a+g^{abc}V_i^b\pi_i^c+
\left( g^{abc}+g^{bac}\right)\,V^b_i\partial_iV^c_0\nonumber\\
&  & -g^{bca}\partial_i\left( V^b_0V^c_i\right) -
g^{cba}\partial_i\left( V^b_0V^c_i\right)+ M_{a}^2 V^a_0 \nonumber\\
&& - g^{abc} g^{a'b'c} V_i^b V^{a'}_0 V_i^{b'} -  4 \, h^{abcd}
V_\mu^b V^c_0 V^{d \mu} \nonumber\\
&\equiv&A^{ab}z^b+\chi^a \approx 0,\quad a=1,2,3. \label{equivphi2}
\end{eqnarray}
   Using Eq.\ (\ref{3couplings}), defining $\gamma_1=g_5+g_7$ and
$\gamma_2=g_4+g_6-2g_2$, the matrix $A$ is given by
\begin{equation}
\label{defA} A=\left(\begin{array}{ccc} 0&-2\gamma_1 V_0^3&
\gamma_2 V_0^1-\gamma_1 V_0^2\\
2\gamma_1V_0^3 & 0 & \gamma_1 V_0^1+\gamma_2 V_0^2\\
-(\gamma_2 V_0^1-\gamma_1 V_0^2)& -(\gamma_1 V_0^1+\gamma_2 V_0^2)&0
\end{array}\right).
\end{equation}
   Since the determinant of $A$ vanishes, the three expressions
on the left-hand side of the equations
\begin{equation}
A^{a b} z^b= -\chi^a \label{equations1}
\end{equation}
are not independent.
   As a result, the system of equations (\ref{equations1}) can be
satisfied only if the right-hand sides satisfy the secondary
constraint
\begin{equation}
\phi_2= \chi^1\,(\gamma_1 V_0^1+\gamma_2 V_0^2) + \chi^2\, (\gamma_1
V_0^2-\gamma_2 V_0^1)-\chi^3\, 2 \gamma_1\, V_0^3 \approx 0\,.
\label{additionalconstraint1}
\end{equation}
   Let us consider Eq.~(\ref{equations1}) for the case where at least
one of $\gamma_1$ or $\gamma_2$ does not vanish.
   For non-vanishing $V_0^1$ and/or $V_0^2$ we obtain
\begin{eqnarray}
z^1 & = & \frac{\chi_3+\gamma_1 z^2\,V_0^1+\gamma_2 \,z^2
V_0^2}{\gamma_1\, V_0^2-\gamma_2\, V_0^1},\nonumber\\
z^3 & = & \frac{\chi_1+2\,
\gamma_1\,z^2\,V_0^3}{\gamma_2\,V_0^1-\gamma_1\,V_0^2}.
\label{zsolutions}
\end{eqnarray}
   Finally, $z^2$ can be solved from the time conservation of the constraint of
Eq.~(\ref{additionalconstraint1}), $\left\{ \phi_2,H_1\right\}
\approx 0$.
   However, this chain leads to the wrong number of constraints
of the second class \cite{second_class} for our system of three
massive vector fields, namely $3+1=4$ rather than $3+3=6$
constraints.
   In other words, for a self-consistent theory we have to require
\begin{equation}
g_7= - g_5\,, \ \ \  2g_2=g_4+g_6\,. \label{gfirstconstraint}
\end{equation}
   In this case none of the $z^b$ can be solved from
Eq.~(\ref{equivphi2}) and
\begin{eqnarray}
\left\{ \phi^a_1,H_1\right\}&=&
\partial_i\pi_i^a+g^{abc}V_i^b\pi_i^c+
\left( g^{abc}+g^{bac}\right)\,V^b_i\partial_iV^c_0\nonumber\\
&  & -g^{bca}\partial_i\left( V^b_0V^c_i\right) -
g^{cba}\partial_i\left( V^b_0V^c_i\right)+ M_{a}^2 V^a_0 \nonumber\\
&& - g^{abc} g^{a'b'c} V_i^b V^{a'}_0 V_i^{b'} -  4 \, h^{abcd}
V_\mu^b V^c_0 V^{d \mu} \nonumber\\
&\equiv& \phi_2^a \approx 0,\quad a=1,2,3,
\label{equivphi22}
\end{eqnarray}
are the secondary constraints.
   They also have to be conserved in time, i.e.\ their Poisson brackets
with the Hamiltonian have to vanish.
   To obtain the right number of degrees of freedom for
massive vector bosons, the $z^a$ have to be solvable from this
condition.
   In such a case no more constraints occur and the
Lagrangian describes the system with constraints of the second
class.

   Taking Eq.~(\ref{gfirstconstraint}) into account,
we obtain a system of three linear equations for the $z^a$,
\begin{equation}
 \left\{ \phi^a_2,H_1\right\} = {\cal M}^{ab} z^b +Y^a \approx
0,\quad a=1,2,3.
\label{zaequations}
\end{equation}
   The $3\times 3$ matrix ${\cal M} $ is given by
\begin{eqnarray}
{\cal M}^{ab} & = & M^2_a\delta^{ab}
-\left(g^{bca}+g^{cba}\right)\partial_i V_i^c\nonumber\\
&& -\left( g^{ace} g^{bde} - 4 h^{acbd} \right) V_i^c V_i^d\nonumber\\
&& - 4 \left( 2 h^{abcd} + h^{acbd}\right) V_0^c V_0^d,
\label{commutmatr}
\end{eqnarray}
and the $Y^a$ are some functions of the fields and conjugated
momenta, the particular form of which plays no role in the
following discussion.
   If the determinant of ${\cal M}$ vanishes for any values of the
fields, then the $z^a$ cannot generally be determined and additional
constraints have to be imposed \cite{gitman}.
   However, then one generates the wrong number of degrees of freedom.
   This problem, in its various appearances, is known as the
Johnson-Sudarshan \cite{Johnson:1960vt} and the Velo-Zwanziger
\cite{Velo:1969bt} problem.
   To obtain a self-consistent field theory we have to demand that
${\rm det} {\cal M}$ does not vanish for any values of the
fields.
   We refrain from displaying the lengthy expression of the
determinant for arbitrary fields.
   In the appendix we provide the analysis leading to
the following conditions for the coupling constants:
\begin{eqnarray}
\label{ccconditions}
&&g_1=g_2=0,\nonumber\\
&&d_2=-d_1,\nonumber\\
&&d_1\geq \frac{g_3^2}{2},\nonumber\\
&&d_4=-d_3,\nonumber\\
&&d_3\leq -g_4^2-g_5^2,\nonumber\\
&&d_5=0.
\end{eqnarray}
   Note that Eqs.~(\ref{3couplings}) together with Eqs.~(\ref{gfirstconstraint}) and
(\ref{ccconditions}) imply
\begin{equation}
\label{gantisymmetricab}
g^{abc}=-g^{bac},
\end{equation}
i.e.~the three-vector vertex couplings are antisymmetric in the first two
indices.

\section{Quantization}
   In the following, we assume that the coupling constants are related
to each other in such a way that ${\rm det}\, {\cal M}$ does not vanish and proceed
with the quantization.
   Note that, for small values of the fields and their derivatives,
the $z^a$ can be solved from Eq.~(\ref{zaequations}) as a perturbative expansion
in the coupling constants.
   According to a general theorem for systems with second-class
constraints \cite{gitman} there always exists a canonical set of dynamical
variables (fields and conjugated momentum fields) with the
following properties: the set of variables may be divided into two
subsets $\{\omega\}$ and $\{\Omega\}$, each consisting of canonically
conjugate pairs such that (a) the constraints
only appear among conjugate pairs of $\{\Omega\}$ and (b) the original
and the new set of constraints are equivalent.
   The $\omega$ are dynamical variables and the dynamics is described by
the physical Hamiltonian
\begin{equation}
{\cal H}^{\rm ph}= {\cal H}(\omega,\Omega)|_{\Omega=0}={\cal
H}^{\rm ph}(\omega)\,.\label{phH}
\end{equation}
   As soon as the dynamical canonical variables and the Hamiltonian
have been identified, the quantization can be performed using the
path integral method.
   Let $\omega^{1}$ and $\omega^{2}$ denote the
canonical fields and the respective conjugate momentum fields, both belonging
to the set $\{\omega\}$.
   The generating functional reads
\begin{equation}
Z[J^\omega] = \int {\cal D} \omega\,e^{i \int d^4 x \,\left[
\omega^{2}\dot\omega^{1}-{\cal H^{\rm ph}}(\omega)+J^\omega \omega
\right]}\,. \label{GFP}
\end{equation}
   By introducing a product of functional delta functions,
$\delta(\Omega)$, involving the constraints in terms of
the set $\{\Omega\}$, Eq.~(\ref{GFP}) is expressed as
\begin{equation}
Z[J^\omega] = \int {\cal D} \omega\,{\cal D}
\Omega\,\delta(\Omega)\,e^{i \int d^4 x \,\left[ \omega^{2
}\dot\omega^{1}+\Omega^{2}\dot\Omega^{1}-{\cal
H}(\omega,\Omega)+J^\omega\,\omega +J^\Omega\,\Omega
\right]}=Z[J]\,, \label{GFPRewritten}
\end{equation}
where $J=(J^\omega,J^\Omega)$, in addition, generates a coupling
to the variables $\Omega^1$ and $\Omega^2$ of $\{\Omega\}$.
   Now we switch to the original variables.
   The functional $\delta$ function in Eq.~(\ref{GFPRewritten}) can be written as
\begin{equation}
\delta(\Omega) = \delta(\phi)\,\left[{\rm det}{\cal \left\{
\phi,\phi\right\}}\right]^{1/2}\,,   \label{deltarewrite}
\end{equation}
where $\phi$ denotes the original system of constraints and
\begin{equation}
\left\{\phi,\phi\right\}_{(al), (bk)}=\left\{\phi^a_l,\phi^b_{k}\right\}
\label{phimatrix}
\end{equation}
is, in our specific case, the $6\times 6$ matrix of the Poisson brackets of constraints
[see Eqs.~(\ref{phi1}) and (\ref{equivphi22})].
   The Poisson brackets $\left\{\phi^a_1,\phi^b_{1}\right\}$ vanish,
once Eq.~(\ref{gfirstconstraint}) is taken into account. As a consequence,
the entries $\left\{\phi^a_2,\phi^b_{2}\right\}$ do not contribute to the
determinant.
   The square root is, thus, of the form
\begin{equation}
\left[{\rm det}{\cal \left\{ \phi,\phi\right\}}\right]^{1/2}={\rm
det} \left\{ \phi_1^a,\phi_2^b\right\}={\rm det} \left\{{\cal
M}^{ab}\right\}, \label{functdet}
\end{equation}
where the $3\times 3$ matrix ${\cal M}$ is given by
Eq.~(\ref{commutmatr}) supplemented with the relations among
coupling constants obtained above.

As $(\omega,\Omega)$ are canonical variables, the
Jacobian corresponding to a change of variables to the original
ones, is equal to one.
   Also the action is a canonically invariant quantity,
\begin{equation}
S=\int d^4 x \left[\omega^{2
}\dot\omega^{1}+\Omega^{2}\dot\Omega^{1}-{\cal
H}(\omega,\Omega)\right]= \int d^4 x\left[\pi_0^{a }\dot
V_0^{a}+\pi_i^{a}\dot V_i^{a}-{\cal H}(V,\pi)\right]\,.
\label{action}
\end{equation}
   In the following, $\{J^{a\mu}\}$ will symbolically denote the
set of external sources coupling to the vector fields of the original Lagrangian.
   After the change of variables, we trade the generating functional
of Eq.\ (\ref{GFPRewritten}) for a generating functional containing
$\{J^{a\mu}\}$ only,
\begin{equation}
Z[\{J^{a\mu}\}] = \int {\cal D} V\,{\cal D} \pi\,\delta(\phi)\, \left[{\rm
det}{\cal \left\{ \phi,\phi\right\}}\right]^{1/2}\,e^{i \int d^4 x
\,\left[\pi_0^{a}\dot V_0^{a}+\pi_i^{a}\dot V_i^{a}-{\cal
H}(V,\pi)+J^{a\mu}V^a_\mu\right]}\,. \label{GFPRewritten2}
\end{equation}
Next we write $\delta (\phi_2)$ and $\left[{\rm det}{\cal \left\{
\phi,\phi\right\}}\right]^{1/2}$ as functional integrals
\begin{eqnarray}
\delta(\phi_2) & \sim & \int {\cal D} \lambda\,e^{i\int d^4 x\,
\lambda^a \phi_2^a } \label{deltafunctintegral}\,,\nonumber\\
\left[{\rm det}{\cal \left\{ \phi,\phi\right\}}\right]^{1/2} &
\sim & \int {\cal D}\,c\,{\cal D}\,\bar c \,e^{i \int d^4 x
\,{\cal L}_{\rm ghost}}\,, \label{determinantghost}
\end{eqnarray}
where the ghost Lagrangian reads
\begin{eqnarray}
{\cal L}_{\rm ghost} & = & M^2_a\,\bar c^a c^a
- \left( g^{ace} g^{bde} - 4 h^{bdac} \right) \, V_i^c
V_i^d \,\bar c^a c^b\nonumber\\
&& - 4 \left(2 h^{bacd} +h^{bcad}\right) V_0^c V_0^d\, \bar c^a
c^b\,.\label{ghostlagrangian}
\end{eqnarray}
By substituting Eqs.~(\ref{determinantghost}) with
(\ref{ghostlagrangian}) in Eq.~(\ref{GFPRewritten2}) and shifting the
integration variable $V_0^a\to V_0^a-\lambda^a$, we obtain
\begin{equation}
Z[\{J^{a\mu}\}] = \int {\cal D} V\,{\cal D} \pi\,{\cal D}\,c\,{\cal D}\,\bar
c\,{\cal D} \lambda\, \delta(\tilde\phi_1)\,e^{i \int d^4 x
\,\left( {\cal K}+J^{a\mu}V_\mu^a\right)}\,,
\label{GFPEffective}
\end{equation}
where
\begin{eqnarray}
\tilde\phi_1^a & = & \pi^a_0+g^{bca} \left(V^b_0-\lambda^b\right)
\left(V^c_0-\lambda^c\right)\,,
\nonumber\\
{\cal K} & = & \pi_0^{a }\dot V_0^{a}+\pi_i^{a}\dot
V_i^{a}-{\cal H}(V,\pi)+\lambda^a\, \phi_2^a+{\cal L}_{\rm
ghost}\nonumber\\
& = & \pi_0^{a }\dot V_0^{a}+\pi_i^{a}\dot
V_i^{a}-{\pi^{a}_i\pi^{a}_i\over 2} -\pi_i^a
\partial_i V^a_0 - \frac{1}{4} V^a_{ij}
V^a_{ij} + {M_a^2\over 2}V^a_\mu V^{a \mu}\nonumber\\
& &+g^{abc}V^a_0 V^b_i \pi^c_i - g^{abc} V_i^a V_j^b
\partial_i V^c_j-\frac{1}{2} g^{abc} g^{a'b'c} V_0^a V_i^b
V^{a'}_0 V_i^{b'} - h^{abcd} V_\mu^a V_\nu^b V^{c \mu} V^{d \nu}\nonumber\\
&& -\pi_0^{a }\dot \lambda^{a} - {M_a^2\over 2}\lambda^a
\lambda^{a} +\frac{1}{2} g^{abc} g^{a'b'c} \lambda^a V_i^b
\lambda^{a'} V_i^{b'} \nonumber\\
& &+ h^{abcd} \left(2\,\lambda^a\lambda^c V_\mu^b V^{d \mu}
+4\,\lambda^c\lambda^d V_0^a
V^{b}_0-8\,V_0^a\lambda^b\lambda^c\lambda^d
+3\,\lambda^a\lambda^b\lambda^c\lambda^d\right)\nonumber\\
& & + M^2_a\,\bar c^a c^a - \left( g^{ace} g^{bde} - 4 h^{bdac}
\right) \, V_i^c
V_i^d \,\bar c^a c^b\nonumber\\
&& - 4 \left( 2 h^{bacd} +h^{bcad}\right) \left(V_0^c
V_0^d-\lambda^c V_0^d-V_0^c \lambda^d+\lambda^c \lambda^d\right)\,
\bar c^a c^b\,. \label{actionrewritten}
\end{eqnarray}
Integrating over $\pi^a_\mu$ we obtain for the generating
functional
\begin{equation}
Z[\{J^{a\mu}\}] = \int {\cal D} V\,{\cal D}\,c\,{\cal D}\,\bar c\,{\cal D}
\lambda\,e^{i \int d^4 x \,\left( {\cal L}_{\rm
eff}+J^{a\mu}V_\mu^a\right)}\,,
\label{GFPEffectiveCanonicalcoordinates}
\end{equation}
where
\begin{eqnarray}
{\cal L}_{\rm eff} & = &
{\cal L}_A - {M_a^2\over 2}\lambda^a \lambda^{a}
+\left(\frac{1}{2} g^{abe} g^{cde}-2 h^{abcd} \right)\lambda^a
V_i^b
\lambda^{c} V_i^{d} \nonumber\\
& &+2 \left( 2 h^{bacd} +h^{bcad}\right) V_0^c
V_0^d\lambda^a\lambda^b
- h^{abcd}
\left(8\,V_0^a\lambda^b\lambda^c\lambda^d
- 3\,\lambda^a\lambda^b\lambda^c\lambda^d\right)\nonumber\\
& & + M^2_a\,\bar c^a c^a - \left( g^{ace} g^{bde} - 4 h^{bdac}
\right) \, V_i^c
V_i^d \,\bar c^a c^b\nonumber\\
&& - 4 \left( 2 h^{bacd} +h^{bcad}\right) \left(V_0^c
V_0^d-\lambda^c V_0^d-V_0^c \lambda^d+\lambda^c \lambda^d\right)\,
\bar c^a c^b\,, \label{effectivelagrangian}
\end{eqnarray}
and the Lagrangian ${\cal L}_A$ is given by
Eqs.~(\ref{LagGeneral}), (\ref{kvLagr}), (\ref{3VintLagr}), and
(\ref{4VL}).

\section{Perturbative renormalizability}

   Before turning to the relations which originate from demanding
perturbative renormalizability, let us recall the constraints which
have been obtained so far from the analysis of the determinant of
${\cal M}$ and the corresponding symmetry input.
   Table \ref{number_of_parameters} contains a summary of the
number of parameters of the unconstrained Lagrangian together with
the number of independent parameters after the constraint analysis.
   Equations~(\ref{ccconditions}) already put severe restrictions on
the coupling constants of the Lagrangian.
   For example, if, in addition, we demand an invariance
under global SU(2) isospin transformations, then the Lagrangian of
Eq.~(\ref{LagGeneral}) contains only four real constants, namely,
one mass parameter $M$, one three-vector-interaction coupling
constant $g$, and two four-vector-interaction coupling constants
$h_1$ and $h_2$:
\begin{equation}
{\cal L}_{\mbox{\tiny SU(2)}}  =  -{1\over 4} \ V^a_{\mu\nu}
V^{a \mu\nu} +\frac{M^2}{2} V_\mu^a V^{a \mu} -g \epsilon^{abc}
V^a_\mu
V^b_\nu \partial^\mu V^{c \nu}  -  h_1 V^a_\mu V^{a\mu} V^b_\nu V^{b\nu} -h_2 V^a_\mu V^a_\nu
V^{b\mu} V^{b\nu}. \label{lsu2}
\end{equation}
   Also note that enforcing isospin symmetry in the present case
entails charge-conjugation invariance.
   Applying the above constraint analysis results in
the two inequalities
\begin{eqnarray}
\label{inequalitiessu2}
h_1&\geq& \frac{g^2}{4},\nonumber\\
h_2&\leq& -\frac{g^2}{4}.
\end{eqnarray}

\begin{table}
\caption{\label{number_of_parameters} Number of parameters before
and after the constraint analysis depending on the symmetry input.
The Lagrangian without internal symmetry contains 51 real parameters,
namely 3 masses ($M$), 27 cubic coupling constants ($g$), and 21 quartic
coupling constants ($h$).
   In all cases, the constraint analysis provides two additional inequalities
   [see Eqs.\ (\ref{ccconditions})].
}
\begin{ruledtabular}
   \begin{tabular}{ccccc}
   Symmetry & Parameters before & Total number & Parameters afterwards & Total number \\
U(1)& 2 $M$, 7 $g$, 5 $h$ & 14 & 2 $M$, 3 $g$, 2 $h$ & 7\\
U(1) + C & 2 $M$, 3 $g$, 5 $h$ & 10 & 2 $M$, 2 $g$, 2 $h$  & 6\\
SU(2) & 1 $M$, 1 $g$, 2 $h$ & 4 & 1 $M$, 1 $g$, 2 $h$ & 4
   \end{tabular}
\end{ruledtabular}
\end{table}

    We will now show that further relations among the coupling constants
result by demanding that the ultraviolet divergences of the loop diagrams
can be absorbed in the redefinition of masses, coupling constants, and fields of
the most general effective Lagrangian containing {\it all} terms
consistent with the assumed underlying symmetries.
   In particular, starting from the classical theory with a U(1) symmetry only,
we will be led to infer the SU(2) symmetry of the quantized theory from perturbative
renormalizability, instead of using it as an {\em input} to the analysis.
   Moreover, the inequalities of Eqs.\ (\ref{inequalitiessu2})
will be replaced $h_1=-h_2=g^2/4$.
   In other words, we will obtain the standard Yang-Mills Lagrangian
plus an additional mass term.

   Let us discuss the logarithmic divergences.
   Although dimensional regularization ignores the power-law divergences,
it keeps track of all logarithmic divergences.
   Because the fields $\lambda^a$, $c^a$, and $\bar c^a$ do not have kinetic
parts in Eq.~(\ref{effectivelagrangian}), their contributions to perturbative
calculations vanish in dimensional regularization.

\begin{figure}
\epsfig{file=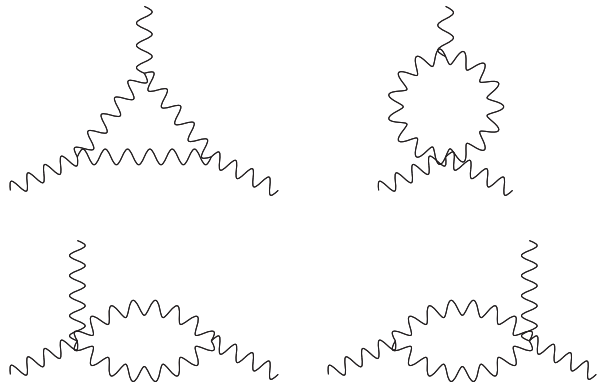, width=8truecm}
\caption[]{\label{3vectorvertexV:fig} One-loop contributions to the
three-vector vertex function. The wiggly line corresponds to the
vector meson.}
\end{figure}

\begin{figure}
\epsfig{file=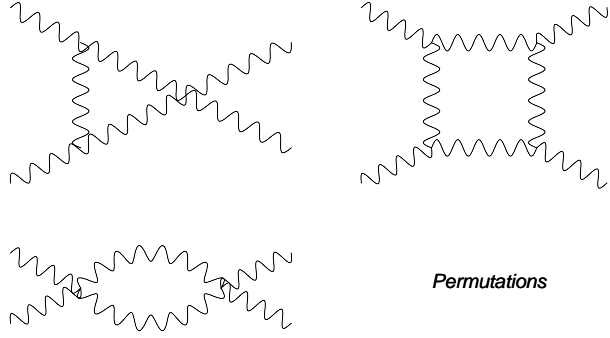, width=8truecm}
\caption[]{\label{4vectorvertexV:fig} One-loop contributions to the
four-vector vertex function. The wiggly line corresponds to the
vector meson.}
\end{figure}

   We start with the logarithmically divergent parts of the
vertex functions involving neutral fields only.
   As $g^{333}=g_1=0$ [see Eqs.~(\ref{ccconditions})], we require
that the divergent part of the $V^3V^3V^3$ vertex function vanishes
(see Fig.~\ref{3vectorvertexV:fig}).
   This leads to
\begin{equation}
g_4 \left(5 g_4^2+3 g_5^2+3
   d_3\right)=0.
\label{condition1}
\end{equation}
   For the $V^3V^3V^3V^3$ vertex function (see Fig.~\ref{4vectorvertexV:fig}),
the same reasoning results in
\begin{equation}
5 g_4^4+2 g_5^2 g_4^2+5
   g_5^4+5 d_3^2+2 d_3
   \left(g_4^2+5 g_5^2\right)=0.
\label{condition2}
\end{equation}
   The unique solution to Eqs.~(\ref{condition1}) and (\ref{condition2}) reads
\begin{equation}
g_4=0\,,\quad d_3 = -g_5^2\,. \label{d3g4}
\end{equation}
   Note that one of the two inequalities of Eqs.~(\ref{ccconditions})
is thereby replaced in terms of an equality.

   Next, we demand that the divergent part of the one-loop
contribution to the $V^1V^2V^3$ vertex function can be absorbed in the
renormalization of the corresponding tree-order vertex.
   From this requirement we obtain
$$
-g_5 \left[\left(M_3^4+2 M_3^2 M^2 \right)
   g_3^2+6 M_3^2 M^2 g_3 g_5
   + \left(M^4+2 M^2 M_3^2\right)g_5^2 \right]=0.
$$
   The solution is either $g_5=0$ (with no constraints for the values
of $g_3$, $M$, and $M_3$) or for $g_5\neq 0$:
\begin{equation}
M_3=M\quad\mbox{and}\quad g_3=-g_5.
\label{eqalternative}
\end{equation}
   As there is no tree-order contribution to the vertex function $V^1V^1V^1V^1$,
we demand that the divergent part of the one-loop contribution vanishes.
   Taking Eq.~(\ref{d3g4}) for $g_4$ and $d_3$ into account, we obtain
\begin{eqnarray}
&& 60 d_1^2 M^4 M_3^4+8 g_3^3 g_5
   M^2 \left(M_3^2-M^2\right) M_3^4+g_3^4
   \left(15 M_3^4-4 M^2 M_3^2+M^4\right) M_3^4 \nonumber\\
&& +4
   g_3^2 g_5^2 M^4
   \left(M_3^2-M^2\right) M_3^2 +4 g_3 g_5^3
   M^4 \left(2 M_3^2+M^2\right) M_3^2+5
   g_5^4 M^4 \left(2 M_3^4+M^4\right) \nonumber\\
&& -4
   d_1 \biggl[2 g_3 g_5 M^2
   \left(M_3^2-4 M^2\right) M_3^4+g_5^2
   M^4 \left(5 M_3^2-2 M^2\right)
   M_3^2\nonumber\\
&& +g_3^2 \left(5 M_3^8-2 M^2 M_3^6+3
   M^4 M_3^4\right)\biggr]=0.
\label{eqx}
\end{eqnarray}
   As a result of Eqs.~(\ref{eqalternative}) and (\ref{eqx}) we obtain that
either
\begin{eqnarray}
d_1 & = & \frac{g_3^2}{2}\,,\nonumber\\
g_5 & = & -g_3\,, \ \ M_3=M \,, \label{d1eq}
\end{eqnarray}
or
\begin{eqnarray}
d_1 & = & \frac{g_3^2}{30 M^4} \Biggl(5 M_3^4-2 M^2 M_3^2+3
M^4\nonumber\\
&&  \pm  \sqrt{25 M_3^8-20 M^2 M_3^6-191 M^4
M_3^4+48 M^6 M_3^2-6 M^8}\Biggr)\,,\nonumber\\
g_5 & = & 0\,. \label{d1eq2}
\end{eqnarray}
   The solution given in Eq.~(\ref{d1eq2}) relates masses and
dimensionless coupling constants.
   It is therefore not compatible with the assumption that parameters
with different dimensions be independent.
   We are thus left with the solution of Eq.~(\ref{d1eq}).

   Defining $g=g_3$, all the relations among the coupling constants and masses can be
summarized as
\begin{eqnarray}
g^{abc} & = & -g\,\epsilon^{abc} \,,\nonumber\\
h^{abcd} & = &
\frac{1}{4}\,g^{abe}g^{cde}\,,\nonumber\\
M_1 & = & M_2=M_3=M \,.\label{ccconditions1}
\end{eqnarray}
   For the couplings in Eq.~(\ref{ccconditions1}) the determinant
${\rm det}{\cal \left\{ \phi,\phi\right\}}$ does not depend on the
fields, i.e.\ the ghost fields completely decouple from the vector-meson
fields.
   The matrix ${\cal M}^{ab}$ in Eq.~(\ref{zaequations})
becomes $M^2 \delta^{ab}$, i.e.\ all $z^a$ are solved for {\it
all}\, field configurations.
   After integrating over the ghost fields in Eq.\ (\ref{GFPEffectiveCanonicalcoordinates}),
the resulting effective Lagrangian can be written in a compact form
\begin{equation}
{\cal L}_{\rm eff}={\cal L}_A=-{1\over 4} \ G^a_{\mu\nu} G^{a
\mu\nu} +\frac{M^2}{2} V_\mu^a V^{a \mu}\,, \label{EffLagr}
\end{equation}
where
\begin{equation}
G^a_{\mu\nu}=V^a_{\mu\nu}-g\epsilon^{abc}V^b_\mu V^c_\nu\,.
\label{gdefinition}
\end{equation}
   The final generating functional of the Green's functions has the
form
\begin{equation}
Z[\{J^{a\mu}\}] = \int {\cal D} V\,e^{i \int d^4 x \,\left( {\cal
L}_A+J^{a\mu}V^a_\mu\right)} \label{GFFinal}
\end{equation}
and results in ''naive'' Feynman rules.
   The remarkable result is that we have ultimately been led to the
standard SU(2) Yang-Mills Lagrangian with an additional mass term.

\section{Conclusions}

    In this work we have considered the interaction terms with dimensionless coupling
constants of the most general parity-conserving and
Lorentz-invariant effective-field-theory Lagrangian of
self-interacting massive charged and neutral vector fields. We
have analyzed the quantization procedure of this system with
second-class constraints using the canonical formalism. By
demanding that the quantized theory is self-consistent in the
sense of constraints and perturbative renormalizability we
obtained relations among the 12 originally available coupling
constants [U(1) symmetry], resulting in a substantial reduction in
the number of independent couplings.
    In practice quantum theories are usually obtained by quantizing
classical theories. However, it is understood that the natural logic is
exactly the other way around, i.e.\ classical theories should be
obtained as limits of quantum theories. Following this logic, only
those classical theories could be treated as self-consistent, which
are obtained from self-consistent quantum theories.
   To be specific, in the present case the considered model reduces
to a charge-conjugation-invariant SU(2) Yang-Mills vector field theory
with an additional mass term.
   This suggests that the massive Yang-Mills effective Lagrangian is
the most general one and therefore could be used in
phenomenological applications.

\acknowledgments

The authors thank Martin Reuter for interesting discussions.
This work was supported by the Deutsche Forschungsgemeinschaft (SFB 443).

\section{Appendix}
   In this appendix we explicitly display suitable field configurations from
which we have inferred the constraints for the coupling constants summarized in
Eqs.\ (\ref{ccconditions}).
   By restricting ourselves to particular field configurations,
the analysis is greatly simplified without spoiling the general argument,
because $\det \mathcal{M}$ should be non-vanishing for arbitrary values of
the fields.
   The entries of the matrix $\mathcal{M}$ are defined in Eq.~(\ref{commutmatr}).
   Note that, in some cases, at a given step we explicitly make use of constraints
which have been obtained in a previous step.
\begin{enumerate}
\item For $V_i^a V_i^b = V_0^a= \partial_i V_i^1= \partial_i V_i^2= 0$, one obtains
\begin{displaymath}
\mbox{det}\,{\cal M}=\left(M^2-2 \,g_2 \,\partial_i V_i^3\right)^2 \left(M_3^2-2 \,g_1
\,\partial_i V_i^3\right)\,.
\end{displaymath}
   This determinant can only be non-vanishing for arbitrary $\partial_i V_i^3$, if
\begin{equation}
g_1=g_2=0\,. \label{g1g2}
\end{equation}
\item
   In combination with Eqs.~(\ref{gfirstconstraint}), we are left with only three
independent $g$ couplings, say, $g_3$, $g_4$, and $g_5$:
\begin{displaymath}
g_1=g_2=0,\quad
g_6=-g_4,\quad g_7=-g_5,
\end{displaymath}
   which, together with Eqs.~(\ref{3couplings}), also implies
Eq.~(\ref{gantisymmetricab}), i.e.~$g^{abc}=-g^{bac}$.
   Therefore, we can omit from now on the term proportional to $\partial_i V_i^c$
in ${\cal M}^{ab}$:
\begin{equation}
{\cal M}^{ab}=M^2_a\delta^{ab}
-\left( g^{ace} g^{bde} - 4 h^{acbd} \right) V_i^c V_i^d
- 4 \left( 2 h^{abcd} + h^{acbd}\right) V_0^c V_0^d.
\label{commutmatrnew}
\end{equation}
   Moreover, using the permutation symmetries of Eq.~(\ref{gabcdsym}), we obtain
${\cal M}^{ab}={\cal M}^{ba}$.
   Thus, the evaluation of the determinant simplifies to
\begin{equation}
\mbox{det}\,{\cal M}={\cal M}^{11}{\cal M}^{22}{\cal M}^{33}
-{\cal M}^{11}\left({\cal M}^{23}\right)^2
-{\cal M}^{22}\left({\cal M}^{13}\right)^2
-{\cal M}^{33}\left({\cal M}^{12}\right)^2
+2{\cal M}^{12}{\cal M}^{13}{\cal M}^{23}.
\end{equation}
\item Next, we will investigate field configurations resulting in a diagonal
matrix ${\cal M}$.
\begin{enumerate}
\item
Let us consider arbitrary $x:=V_i^1 V_i^1\geq 0$ and $y:=V_0^1 V_0^1\geq 0$, and set
all the remaining fields to zero. We than have
\begin{eqnarray*}
{\cal M}^{11}&=&M^2+(d_1+d_2)x-3(d_1+d_2)y,\\
{\cal M}^{22}&=&M^2-(g_3^2-d_1+d_2)x-(d_1+d_2)y,\\
{\cal M}^{33}&=&M_3^2-(g_4^2+g_5^2-d_4)x-(d_3+d_4)y.
\end{eqnarray*}
   From ${\cal M}^{11}$, we infer for $x=0$ and arbitrary $y$ the inequality
$d_1+d_2\leq 0$ and for $y=0$ and arbitrary $x$ the
inequality $d_1+d_2\geq 0$.
   Both results combine into
\begin{equation}
d_2=-d_1.
\label{d2d1}
\end{equation}
   Using this result we infer from ${\cal M}^{22}$ for arbitrary $x$ the
inequality
\begin{equation}
\label{ineqdg3}
d_1\geq\frac{g_3^2}{2}.
\end{equation}
   Similarly, from ${\cal M}^{33}$ we infer the inequalities
\begin{eqnarray}
d_3+d_4&\leq& 0,\\
g_4^2+g_5^2&\leq&d_4.
\end{eqnarray}
\item Let us consider arbitrary $x:=V_i^3 V_i^3\geq 0$ and $y:=V_0^3 V_0^3\geq 0$, and set
all the remaining fields to zero. We than have
\begin{eqnarray*}
{\cal M}^{11}&=&M^2-(g_4^2+g_5^2-d_4)x-(d_3+d_4)y={\cal M}^{22},\\
{\cal M}^{33}&=&M_3^2+4d_5x-12 d_5 y.
\end{eqnarray*}
   While ${\cal M}^{11}$ does not provide any new information, we infer
from ${\cal M}^{33}$ the condition
\begin{equation}
d_5=0.
\label{d50}
\end{equation}
\end{enumerate}
\item We now turn to a configuration involving also off-diagonal elements of
${\cal M}$. Let us consider $x:=V_0^2 V_0^2=V_0^3 V_0^3=V_0^2 V_0^3\geq 0$
and set all the remaining fields to zero. We then obtain
\begin{eqnarray*}
\mbox{det}\,{\cal M}&=&{\cal M}^{11}[{\cal M}^{22}{\cal M}^{33}-({\cal M}^{23})^2]\\
&=&[M^2-(d_3+d_4)x]
\{[M^2-(d_3+d_4)x][M^2_3-(d_3+d_4)x]-4(d_3+d_4)^2x^2\}.
\end{eqnarray*}
   We already know that $d_3+d_4\leq 0$.
   The term in the curly braces has a root for a sufficiently large value of $x$,
unless the condition
\begin{equation}
d_4=-d_3\, \label{d42}
\end{equation}
holds.
\item With the constraint analysis performed so far, we end up with
\begin{eqnarray*}
{\cal M}^{11}&=&M^2+\alpha V_i^2 V_i^2+\beta V_i^3 V_i^3,\\
{\cal M}^{22}&=&M^2+\alpha V_i^1 V_i^1+\beta V_i^3 V_i^3,\\
{\cal M}^{33}&=&M^2_3+\beta(V_i^1 V_i^1+V_i^2 V_i^2),\\
{\cal M}^{12}&=&-\alpha V_i^1 V_i^2,\\
{\cal M}^{13}&=&-\beta V_i^1 V_i^3,\\
{\cal M}^{23}&=&-\beta V_i^2 V_i^3,
\end{eqnarray*}
where
\begin{eqnarray*}
\alpha&:=& 2 d_1-g_3^2\geq 0,\\
\beta&:=& -d_3-g_4^2-g_5^2\geq 0.
\end{eqnarray*}
   In particular, all the terms involving the fields $V_0^a$ have disappeared
from ${\cal M}$.
   We finally want to verify that we have reached a point, beyond which we cannot
obtain any further constraints from the analysis of the determinant.
   We will show that, for any field configuration,
$\mbox{det}\,{\cal M}\geq M^4 M_3^2$ as long as $\alpha\geq 0$ and $\beta\geq 0$.
   For that purpose we introduce the abbreviations
\begin{eqnarray*}
u&:=& V_i^1 V_i^1\geq 0,\\
v&:=& V_i^2 V_i^2\geq 0,\\
w&:=& V_i^3 V_i^3\geq 0,\\
x&:=& V_i^1 V_i^3,\\
y&:=& V_i^2 V_i^3,\\
z&:=& V_i^1 V_i^2,
\end{eqnarray*}
in terms of which the determinant reads
\begin{eqnarray*}
\mbox{det}\,{\cal M}&=&
(M^2+\alpha v+\beta w)(M^2+\alpha u+\beta w)[M_3^2+\beta(u+v)]\\
&&-(M^2+\alpha v+\beta w)\beta^2 y^2
-(M^2+\alpha u+\beta w)\beta^2 x^2
-[M^2_3+\beta (u+v)]\alpha^2 z^2\\
&&-2\alpha\beta^2xyz.
\end{eqnarray*}
Making use of $x^2\leq uw$, $y^2\leq vw$, $z^2\leq uv$, and $xyz\leq uvw$, we
obtain as a lower bound for $\mbox{det}\,{\cal M}$,
\begin{eqnarray*}
\mbox{det}\,{\cal M}&\geq&(M^2+\alpha v+\beta w)(M^2+\alpha u+\beta w)
[M_3^2+\beta(u+v)]\\
&&-(M^2+\alpha v+\beta w)\beta^2 vw
-(M^2+\alpha u+\beta w)\beta^2 uw
-[M^2_3+\beta (u+v)]\alpha^2 uv\\
&&-2\alpha\beta^2uvw\\
&=&M^4 M_3^2+M^4\beta(u+v)+M^2 M_3^2[\alpha(u+v)+2\beta w]\\
&&+M^2\beta(u+v)[\alpha(u+v)+\beta w]
+M_3^2\beta w[\alpha(u+v)+\beta w]\\
&\geq& M^4 M_3^2.
\end{eqnarray*}
This completes the analysis leading to the constraints given in
Eqs.\ (\ref{ccconditions}).

\end{enumerate}

%\end{references}


\begin{thebibliography}{99}
\bibitem{3vc}
Structures of the type $\partial^\mu V^a_\mu V^b_\nu V^{c\nu}$ need
not be included, because they are, up to a total derivative, equal
to those given in Eq.~(\ref{3VintLagr}). Moreover, structures of the
type $\epsilon^{\mu\nu\alpha\beta}
\partial_\mu V^a_\nu V^b_\alpha V^c_\beta$ are odd under parity.
\end{thebibliography}
\end{document}